\documentclass[iop]{emulateapj}
\usepackage{apjfonts}
\usepackage{psfig}
\usepackage{mathrsfs}
\usepackage{amsmath}
\usepackage{natbib}
\begin{document}

\slugcomment{AJ, in press}
\shorttitle{NGC~6402. I. CMD}
\shortauthors{C. Contreras Pe\~{n}a et al.}

\title{The Globular Cluster NGC~6402 (M14). I. A New {\it{BV}} Color-Magnitude Diagram\footnote{Based on observations obtained with the 0.9m telescope at the Cerro Tololo Interamerican Observatory, Chile, operated by the SMARTS consortium.}}

\author{C. Contreras Pe\~{n}a,\altaffilmark{1,2,3,4} M. Catelan,\altaffilmark{1,4} F. Grundahl,\altaffilmark{2}
        A. W. Stephens,\altaffilmark{5} H. A. Smith\altaffilmark{6}}

\altaffiltext{1}{Pontificia Universidad Cat\'olica de Chile, Departamento de 
       Astronom\'\i a y Astrof\'\i sica, Av. Vicu\~{n}a Mackenna 4860, 
       782-0436 Macul, Santiago, Chile; e-mail: mcatelan@astro.puc.cl}

\altaffiltext{2}{Danish AsteroSeismology Centre (DASC), Department of Physics and Astronomy, Aarhus University, DK-8000 Aarhus C, Denmark}

\altaffiltext{3}{Centre for Astrophysics Research, Science and Technology Research Institute, University of Hertfordshire, Hatfield, AL10 9AB; e-mail: c.contreras@herts.ac.uk}

\altaffiltext{4}{The Milky Way Millennium Nucleus, Av. Vicu\~{n}a Mackenna 4860, 782-0436 Macul, Santiago, Chile}

\altaffiltext{5}{Gemini Observatory, 670 North A'ohoku Place, Hilo, HI 96720, USA}

\altaffiltext{6}{Department of Physics and Astronomy, Michigan State University, East Lansing, MI 48824}

\begin{abstract}
We present $BV$ photometry of the Galactic globular cluster \objectname{NGC~6402} (M14), based on  65 $V$ frames and 67 $B$ frames, reaching two magnitudes below the turn-off level. This represents, to the best of our knowledge, the deepest color-magnitude diagram (CMD) of  \objectname{NGC~6402} available in the literature. Statistical decontamination of field stars as well as differential reddening corrections are performed in order to derive a precise ridgeline and derive physical parameters of the cluster therefrom. 

We discuss previous attempts to derive a reddening value for the cluster, and argue in favor of a value $E(\bv) = 0.57\pm 0.02$, which is significantly higher than indicated by either the \citeauthor{1982BHM} or \citeauthor{1998SchM} (corrected according to \citeauthor{bonif1}) interstellar dust maps. Differential reddening across the face of the cluster, which we find to be present at the level of $\Delta E(\bv) \approx 0.17$~mag, is taken into account in our analysis. We measure several metallicity indicators based on the position of the red giant branch (RGB) in the cluster CMD. These give a metallicity of ${\rm [Fe/H]} = -1.38 \pm 0.07$ in the \citeauthor{zw84} scale and ${\rm [Fe/H]} = -1.28\pm 0.08$ in the new \citeauthor{cg2} (UVES) scale. We also provide measurements of other important photometric parameters for this cluster, including the position of the RGB luminosity function ``bump'' and the horizontal branch (HB) morphology. We compare the \objectname{NGC~6402} ridgeline with the one for \objectname{NGC~5904 (M5)} derived by \citeauthor{1996Sandquist}, and find evidence that \objectname{NGC~6402} and \objectname{M5} have approximately the same age, to within the uncertainties~-- although the possibility that M14 may be slighlty older cannot be ruled out.
   
\end{abstract}

\keywords{stars: Hertzsprung-Russell diagram --- stars: variables: other --- Galaxy: globular clusters: individual (NGC~6402) --- galaxies: dwarf --- galaxies: star clusters}

\section{Introduction}\label{sec:intro}

NGC 6402 (M14, $\alpha$ = 17:37:36.1, $\delta$ = -03:14:45, J2000) is a moderately high-metallicity globular cluster (GC), with ${\rm [Fe/H]} = -1.28$. It is located 9.3~kpc away from the Sun and 4.0~kpc from the Galactic center. M14 has core and half-light radii of $0.79\arcmin$ and $1.30\arcmin$, respectively. Being located fairly closely to the Galactic plane with $\ell = 21.32\degr$, $b = 14.81\degr$, it is also fairly highly reddened, with $E(\bv)=0.6$. Its fairly high reddening properly explains why, in spite of M14's being the 10th brightest Galactic GC, with $M_{V} = -9.1$, it has been quite neglected in the literature.\footnote{Unless otherwise stated, all of the quoted  information was taken from the 2010 edition of the \citet{h96,wh10} catalog.} 

Indeed, very few color-magnitude diagrams (CMDs) of the cluster have been obtained in the past. To our knowledge, there are only five published CMDs. The first two, by \citet{mir1} and \citet{kogon1}, were only capable of reaching (barely) the horizontal branch (HB) level, being severely dominated by photometric errors. In spite of these limitations, \citeauthor{kogon1} drew attention to the lack of HB stars to the red of the instability strip, despite what is expected for such a high-metallicity cluster. Such a lack of red HB stars is also apparent in the CMDs of \citet{sharam14} and \citet{cote1997}. Instead, a long extension of the blue HB of the cluster also became evident from these studies, particularly the latter. The (snapshot) HST photometry of \citet{piotto2002} revealed more clearly the unusual HB morphology of M14, with stars extending through the RR Lyrae gap to an extended blue tail reaching the main sequence (MS) turnoff level. M14 is thus a prime example of a second-parameter cluster \citep[see, e.g.,][for a recent review and extensive references]{2009Catelan}, in the sense that it has too blue an HB morphology for its metallicity. Yet, the cluster is not included in any of the recent, extensive studies of GC ages \citep[e.g.,][]{area99,sw02,fdaea05,2009Marin-franch,adea10,dvdbea13}. 

Additional interest in M14 is provided by the fact that it is an object which has a suspected extragalactic origin \citep[e.g.,][]{2007Gao}. Several properties of M14 support this possibility. GCs with long blue HB tails, such as M14, have been found to differ in mass and kinematics with respect to GCs without such blue HB tails, possibly pointing to an extragalactic origin for the former \citep{2007Lee}. In addition, M14 shares with $\omega$ Cen (NGC~5139) the peculiar characteristic of possessing a CH star \citep{cote1997}: such stars are more commonly found in dwarf spheroidal galaxies, and indeed $\omega$~Cen has been suggested to be the (present or former) nucleus of one such dwarf \citep[see, e.g.,][]{2010Carretta}. In addition, based on the properties of its RR Lyrae stars, M14 is also found to be an Oosterhoff-intermediate GC (Contreras et al. 2013, {\em in preparation}, hereinafter Paper~II), which is unusual for Galactic GCs, but not so for nearby extragalactic GCs \citep[e.g.,][]{2009Catelan}. A more detailed discussion of the possible extragalactic origin of the cluster will be presented in Paper II. 

This is the latest in our series of papers dealing with the CMDs and variable star content in Galactic GCs \citep[e.g.,][]{2010Contreras,mz1,mz2}. Here we present a deep CMD for the cluster M14, reaching about 2~mag deeper than the MS turn-off. In Contreras et al. (2013) we will describe our comprehensive analysis of the variable star content in the cluster.

\section{Observations and Data Reduction} \label{sec:obs}

Time-series $B$, $V$, and $I$ photometry  was obtained with the 0.9m telescope at CTIO during 16 nights between June 22-July 12 2007, comprising a total of 67, 65, and 62 images in $B$, $V$, and $I$, respectively. Typical exposure times of 600s in $B$, $V$ and 150s in $I$ were used, and the typical seeing was 1.8\arcsec. A 2048$\times$2046 CCD detector with a pixel scale of 0.369\arcsec/pixel, equivalent to a field size of $13.5\arcmin\times 13.5\arcmin$, was used for image acquisition. The detector has a readout noise and gain of 2.7$e^{-}$ and 0.6$e^{-}$/ADU, respectively. The $I$-band data were used only to supplement the variability study that is described in Paper II.

PSF photometry of the images was performed using DAOPHOT~II/ALLFRAME \citep{1994Stetson}. For calibration purposes, five different \citet{1992Landolt} standard fields were observed, comprising a total of 39 standard field images in each filter, giving a total of 220 standard star observations. The color range spanned from $-0.24$ to $2.52$ in \bv, and the airmass $X$ varied from 1.07 to 2.27, with a seeing range of $1\arcsec-2.2\arcsec$. Standard IRAF\footnote{IRAF is distributed by the National Optical Astronomy Observatory, which is operated by the Association of Universities for Research in Astronomy (AURA) under cooperative agreement with the National Science Foundation.} routines were used to produce the final calibrations for $B$, $V$, and $I$ filters. The derived calibration equations are as follows:

\begin{eqnarray}
b &= B & +1.435(\pm0.011) +0.090(\pm0.008)(B-V) \nonumber\\
  &                     & +0.234(\pm0.005)X,\\
v &= V & +1.275(\pm0.009)-0.022(\pm0.006)(B-V)  \nonumber\\
  &                     & +0.118(\pm0.004)X,\\
i &= I & +2.152(\pm0.010)-0.018(\pm0.006)(V-I)  \nonumber\\
  &                     & +0.023(\pm0.004)X\\
\end{eqnarray} 

\noindent (where lower-case letters indicate instrumental magnitudes and $X$ is the airmass), with an rms scatter of 0.036, 0.031, and 0.028~mag for the $B$, $V$, and $I$ solutions, respectively.

%666 stars in V and 656 in B 

\section{Color-Magnitude Diagram} \label{sec:CMD}

\begin{figure}[t]
\begin{center}
\includegraphics[width=8.5cm]{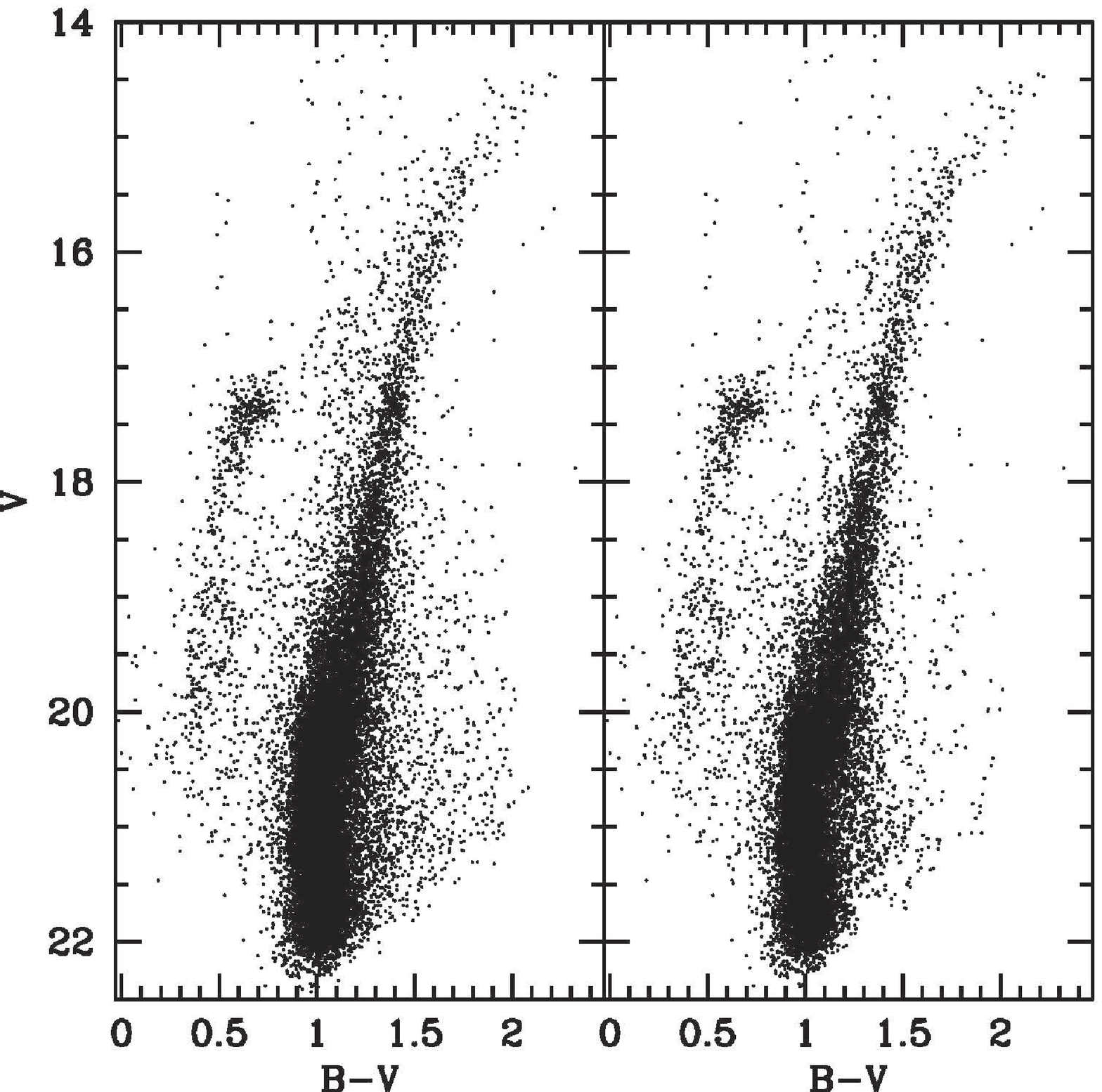}
%\vskip10pt
\includegraphics[width=8.5cm]{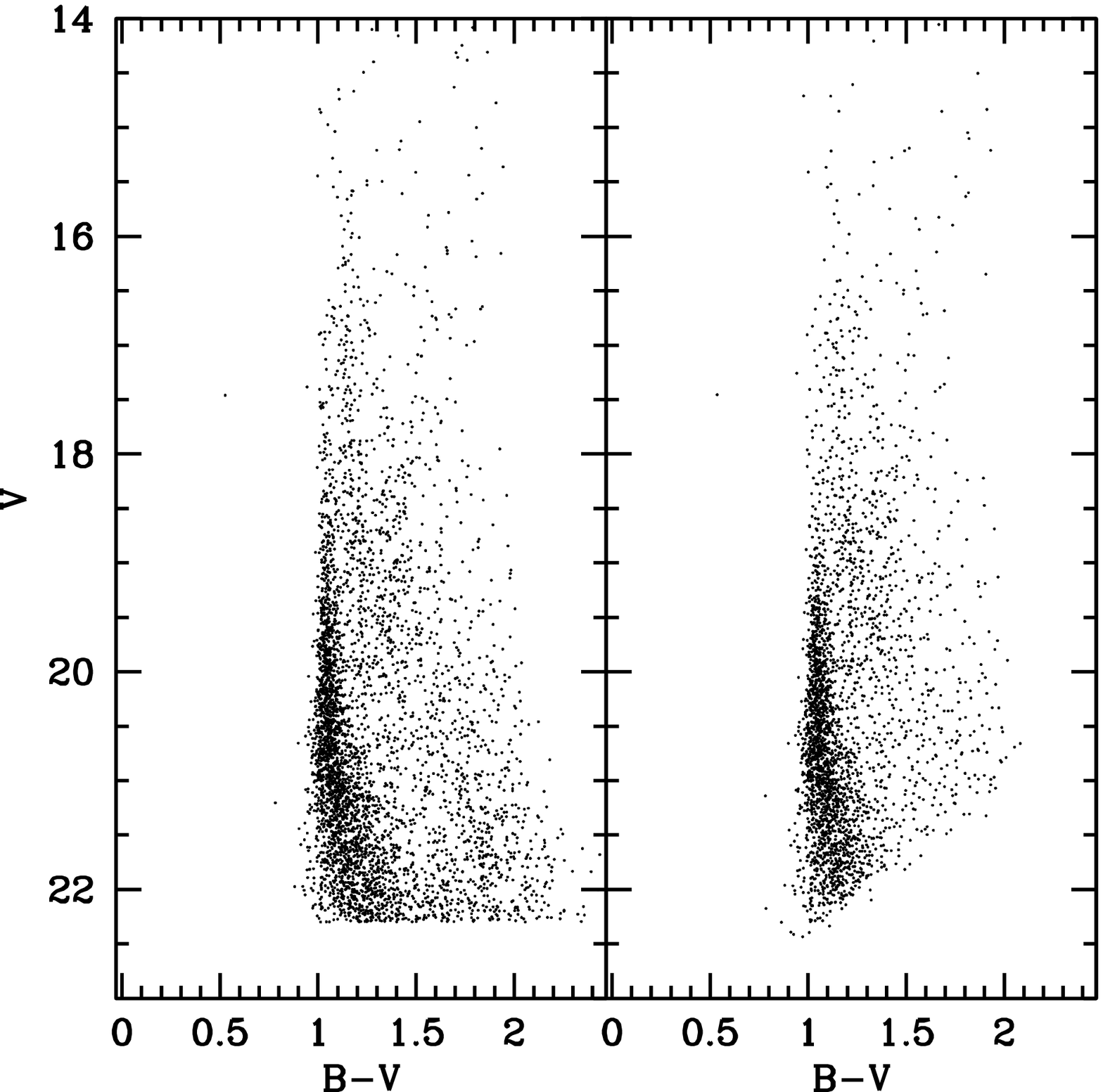}
 \caption{Comparison between CMDs in the field decontamination process.
 {\em Top}: CMD of the cluster before (\textit{left}) and after (\textit{right}) field decontamination. 
 {\em Bottom}: CMD of the field  from Besan\c{c}on models ({\em left}) and that of the subtracted stars ({\em right}). 
 }
\label{fig:decomp}
\end{center}
\end{figure}

In Figure~\ref{fig:decomp} we show the resulting CMD from the DAOPHOT~II/ALLFRAME photometry, after removing stars lying within the innermost regions of the cluster ($r\leq0.86\arcmin$), which are badly affected by crowding. In addition, known variable stars (see Paper II) were also removed from the CMD.

In order to estimate the physical parameters of the cluster from its CMD, it is necessary to define with high precision the cluster ridgeline. However, it is evident from Figure~\ref{fig:decomp} (upper panels) that there is a significant amount of contamination by field stars. In order to account for this effect, we perform statistical decontamination of the observed CMD following the method described in \citet{gal1}, with a maximum distance parameter of $d=0.42$. This method requires the use of a control field; ideally we would use stars lying outside the tidal radius \citep[$r_t = 8.0\arcmin$;][]{wh10} of the cluster. Unfortunately, given the small field of view of the telescope compared to the area of M14, this was not possible in practice. For this reason, the control field is defined by using the Besan\c{c}on Galactic models \citep{bes1}, using as input the relevant parameters for M14, i.e. distance, $\ell$, $b$, projected area in the sky, and reddening~-- with the latter defined as $E(\bv)=0.57$ (see \S\ref{sec:red}).  We note that in the method of \citet{gal1} both the ratio between cluster and control field areas and completeness need to be taken into account. As to the first effect, we note that the area of the Besan\c{c}on control field was chosen to be the same as the area covered by the cluster in the CCD. The second effect becomes important at $V\geq21$~mag, i.e the magnitude at which the number of star per magnitude bin started to drop, thus not affecting the general shape of the cluster's CMD~-- and was thus ignored. 

In Figure~\ref{fig:decomp} we present the result of the statistical decontamination. Here it is possible to observe that some branches, particularly the RGB and the HB, reveal some spread which is larger than would be expected from photometric errors alone. This is especially evident towards the upper part of the CMD. This could be caused by the presence of differential reddening, of the order of several millimagnitudes, across the face of the cluster~-- which would not be unusual, given the high level of extinction in the direction of M14. 

\begin{figure}[t]                            
      \resizebox{8cm}{!}{\includegraphics{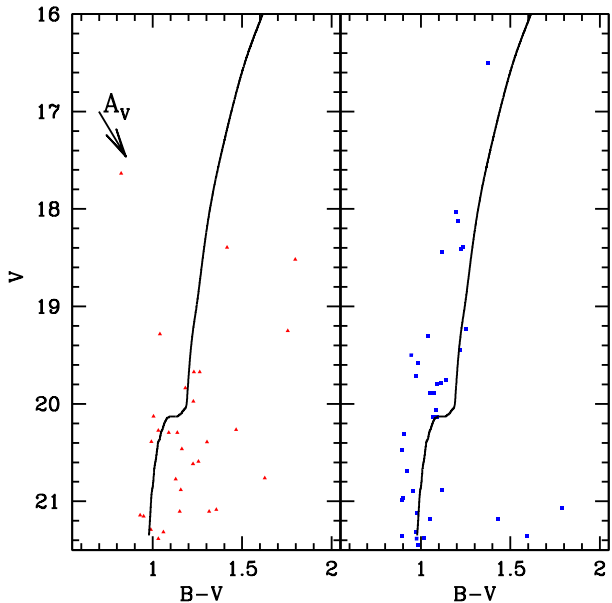}}
      \caption{Illustration of the presence of differential reddening along the face of M14. Blue and red points indicate areas of the cluster with different levels of extinction, compared with the adopted reference area~-- whose CMD is represented by the indicated ridgeline. The reddening vector is schematically indicated in the left panel.}  \label{dred} \end{figure}

To correct for the effect of differential reddening, we followed \citet{piotto1}, dividing the image in one hundred $200\times200$~pix$^{2}$ subsections and constructing a CMD for each. We arbitrarily selected one of the sections as the reference and extracted its fiducial points by drawing a line by hand. Then, a spline function was fitted to the fiducial points thus obtained, in order to obtain a continuous reference line. Figure~\ref{dred} illustrates the detected shifts with respect to this reference line, of stars belonging to two of these small subsections. The reddening vector is also shown in this plot, for clarity. 

\begin{figure}[t]                            
      \resizebox{8cm}{!}{\includegraphics{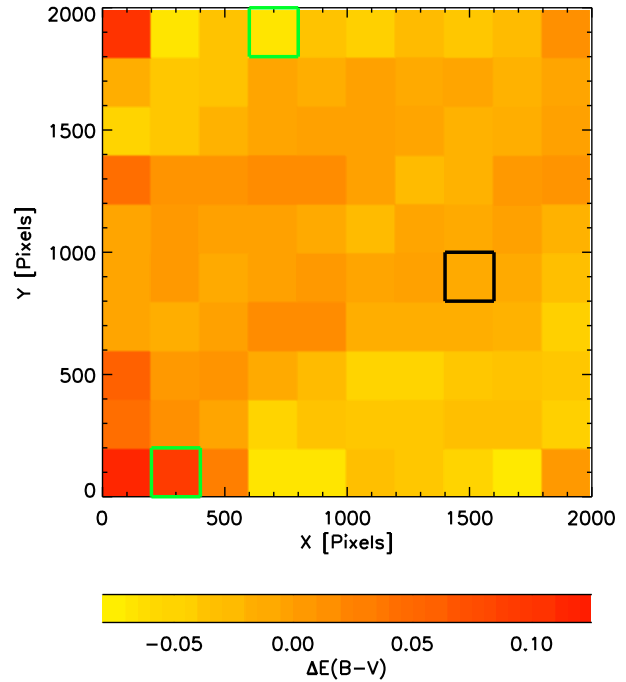}}
      \caption{Reddening map of the cluster, as obtained following the method described in \S\ref{sec:CMD}. The solid black lines mark the reference area, whilst green solid lines mark the areas chosen to illustrate the presence of differential reddening in Figure~\ref{dred}. }  \label{dred_map} \end{figure}

For each subsection, we determine the distance, along the reddening vector [defined by $A_V=3.1 E(\bv)$], between each star with $(\bv) > 0.8$ and $V < 21$ and the reference fiducial CMD ridgeline. Then, $E(\bv)$ for each star is represented by the abscissa of the distance vector, with the total relative reddening in the corresponding subsection defined as the median value of the abscissa in all of the distances. The resulting differential reddening map is presented in Figure~\ref{dred_map}, implying a total range in reddening at the level of $\Delta E(\bv) \approx 0.17$~mag. Every star is then corrected for this differential reddening effect based on its position in the image.  

\begin{figure}[t]                            
      \resizebox{8cm}{!}{\includegraphics{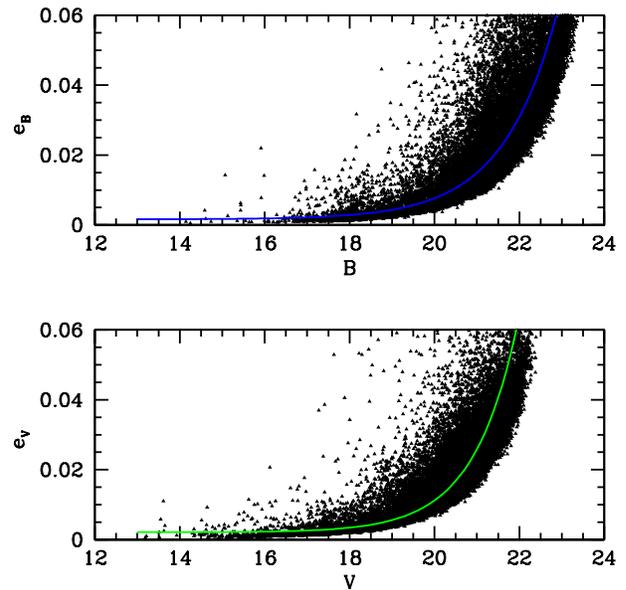}}
      \caption{Photometric errors in $B$ ({\em top}) and $V$ ({\em bottom}) as a function of position in the CMD. The lines indicate exponential fits to the data.}  \label{e_cut} \end{figure}

To further clean our CMD, stars with too large errors compared to what would be expected at their magnitude levels, in both $B$ and $V$, were excluded. An exponential function was fitted to the median values of the error at each magnitude level (Fig.~\ref{e_cut}), and then all stars with errors falling above the fitted line were excluded. This leads to our final CMD for the cluster, which is shown in Figure~\ref{or_comp}. The corresponding photometry is given in Table~\ref{tab:photo}.

\begin{deluxetable*}{llllllll}
\tablewidth{0pc} 
%\tabletypesize{\scriptsize}
\tablecaption{$BV$ Photometry for M14\tablenotemark{a}}
\tablehead{
\colhead{Star} &
\colhead{$\alpha$} &
\colhead{$\delta$} & 
\colhead{$B$} & 
\colhead{$\sigma_B$} & 
\colhead{$V$} &
\colhead{$\sigma_V$} &
\colhead{$\Delta E(B-V)$} 
}  
\startdata
ID-0001 & 17:37:14.32 & -03:11:40.41 & 22.743 & 0.041 & 21.853 & 0.047 &  -0.041 \\
ID-0002 & 17:37:14.56 & -03:14:10.73 & 22.133 & 0.030 & 20.928 & 0.021 &  -0.032 \\
ID-0003 & 17:37:14.57 & -03:18:27.63 & 22.026 & 0.029 & 21.091 & 0.025 &  -0.001 \\
ID-0004 & 17:37:14.60 & -03:12:12.79 & 20.767 & 0.009 & 19.552 & 0.007 &  -0.050 \\
ID-0005 & 17:37:14.60 & -03:15:12.44 & 21.517 & 0.019 & 20.547 & 0.014 &  -0.022 \\
ID-0006 & 17:37:14.63 & -03:18:14.16 & 22.607 & 0.043 & 21.399 & 0.033 &  -0.001 \\
ID-0007 & 17:37:14.71 & -03:12:32.83 & 21.679 & 0.020 & 20.690 & 0.015 &  -0.050 \\
ID-0008 & 17:37:14.79 & -03:14:53.64 & 22.481 & 0.035 & 21.484 & 0.031 &  -0.022 \\
ID-0009 & 17:37:14.82 & -03:13:38.02 & 22.748 & 0.043 & 21.703 & 0.038 &  -0.032 \\
ID-0010 & 17:37:14.84 & -03:15:47.59 & 22.601 & 0.038 & 21.620 & 0.041 &  -0.022 \\
ID-0011 & 17:37:15.30 & -03:12:54.26 & 23.065 & 0.058 & 21.954 & 0.046 &  -0.050 \\
ID-0012 & 17:37:15.73 & -03:09:32.98 & 22.582 & 0.038 & 21.705 & 0.036 &  -0.048 \\
\enddata
\tablenotetext{a}{The complete table is provided in the online version of the journal. The first few rows are provided here for guidance with respect to its contents only.}
\label{tab:photo}
\end{deluxetable*}

\subsection{Mean Reddening}\label{sec:red}
 
The result for metallicity strongly depends on the adopted reddening for the cluster. In this section, we discuss the information regarding the reddening estimate, $E(\bv)$, for M14, as provided in the literature. Before continuing, however, we note that \citet{1978BH} list, in their Table~1, $E(\bv)=0.37$ for M14. However, this is taken from a previous publication by \citet{lee1}, where the author studied the properties of M4 (NGC~6121), not M14. Thus, we will not consider this value further in our analysis.

\subsubsection{Dust Maps}\label{sec:dust}

Information from three all-sky Galactic extinction maps are found in the literature. \citet{2005Amores} give $E(\bv)=0.4$ for their axisymmetric gas distribution model, and $E(\bv)=0.38$ for their spiral model, which takes into account the spiral structure of the Galaxy. \citet{1982BHM} maps give an estimate of $E(\bv)=0.32$ for M14, whereas the dust maps from \citet[][SFD98]{1998SchM} give $E(\bv)=0.48$. However, there are some indications that the latter may overestimate $E(\bv)$ for high reddening values \citep[e.g.,][]{bonif1,2005Amores}. In particular, \citeauthor{bonif1} recommend applying a correction to the SFD98 values, for $E(\bv)_{\rm SFD98} > 0.1$, following the expression $E(\bv)_{\rm SFD98}^{\rm corr} = 0.10+0.65[E(\bv)_{\rm SFD98} -0.10]$ (their Eq.~1). This leads to a corrected reddening value of $E(\bv)_{\rm SFD98}^{\rm corr}=0.347$, which is much closer to the \citeauthor{1982BHM} and \citeauthor{2005Amores} values. We note, in this sense, following \citet{1999Wallick}, that dust maps may indeed suffer from systematic errors. SFD98 maps are based on IRAS/DIRBE measurements of diffuse infrared (IR) emission. Therefore, the presence of cold dust, which is invisible to IRAS, could affect their extinction estimate. The older \citeauthor{1982BHM} estimates, on the other hand, are based on the 21-cm column density and faint galaxy counts, and thus any variable dust-to-gas ratio or galaxy count fluctuations would also affect their $E(\bv)$ value. Accordingly, \citet{h96,wh10} does not take into account the results of extinction maps in the estimation of $E(\bv)$ for any of the clusters listed in his catalog. 

\subsubsection{Photometric Studies}  

\citet{kogon1}, comparing the blue edge of the RR Lyrae instability strip to that of other GCs, and by using the method of \citet{sturch1} of comparing the colors of RR Lyrae stars at minimum light, obtained $E(\bv)=0.35$. However, inspection of the CMD published in their work reveals that the HB is close to the faint end of their photometry, which is badly affected by photometric errors, thus also affecting their estimation of $E(\bv)$. In fact, the assigned error for their reddening estimate is 0.17~mag. \citeauthor{kogon1} also list the metallicity indicator $\Delta V_{1.4}$ to be 1.9~mag; using Equation~3 in \citet{f99} results in a metallicity of ${\rm [Fe/H]}=-0.92$ in the \citet{zw84} scale, which is too high compared to the currently accepted value for M14 \citep[][see \S\ref{sec:intro}]{h96,wh10}. 

\citet{mir1} lists two possible values, namely $E(\bv)=0.63$ (from the blue end of the RR Lyrae region) and $E(\bv) = 0.49$ (from the morphology of his CMD for the cluster). These values agree, within the errors, with those listed by  \citet{kogon1}. However, as in the case of \citeauthor{kogon1}, \citeauthor{mir1}'s results are also strongly affected by the limit in their photometry, and so we decided not to use the values from either of these early studies in our final estimation. 

\subsubsection{Integrated Color-Spectral Type Relation}

\citet{h96,wh10} lists $E(\bv)=0.6$, which is a simple mean of the values listed in \citet{1985Webbink}, namely $E(\bv)=0.58$ \citep[from][]{racine1}; \citet{reed1}, $E(\bv)=0.63$; and \citet{1985Zinn}, $E(\bv)=0.59$ (based on the cluster's integrated $UBV$ colors). 

The relatively high reddening estimate coming from \citet{reed1} seems to be incorrect. \citeauthor{reed1}'s $E(\bv)$ value comes from the relation between the intrinsic color of GCs and their respective spectral types, taken from \citet{1985Hesser}. However, M14 does not follow the tight relation found for most of the clusters in the \citeauthor{reed1} study, with the authors suggesting that Mironov's (1973) estimate should be used, rather than the one found in their own work. 

Indeed, based on the \citet{1985Hesser} study, we see that a spectral type F4 is found for M14.\footnote{The \citet{h96,wh10} catalog also lists this spectral type for M14, based on \citet{1985Hesser}.} However, according to these authors, such a spectral type appears to be too early, the same being noted also by \citet{reed1}. Other methods \citep[e.g.,][]{1959Kinman} classify M14 as having a spectral type F7/8. Taking such a revised spectral type into account in the \citeauthor{reed1} calibration, one finds $E(\bv)=0.55$ for M14. 

Given the uncertainties in the all-sky Galactic extinction maps and of the previous photometric studies of M14, we follow \citet{h96,wh10} and estimate the reddening of M14 by taking a straight average over the values found in \citet{1985Webbink}, \citet{1985Zinn}, and \citet{reed1} (after suitably revising the cluster's spectral type, as just noted). Finally, we obtain $E(\bv)=0.57\pm 0.02$. We again note that this is significantly higher than the values derived on the basis of reddening maps (see \S\ref{sec:dust}). 

\subsection{Metallicity and Reddening}\label{met_sec}

\begin{figure}[t]
\begin{center}
 \resizebox{8cm}{!}{\includegraphics{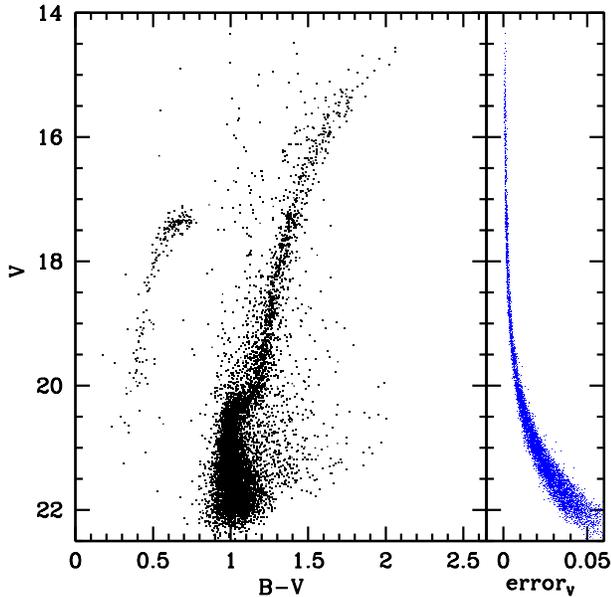}}
 \caption{The M14 CMD ({\em left}), along with the respective errors in $V$ ({\em right}), after statistical field-star decontamination, correction for differential extinction, and selection by photometric errors and radial range.}
 \label{or_comp}
 \end{center}
\end{figure}

In Figure~\ref{or_comp} we show the $B,V$ CMD of M14, after performing the corrections for field contamination and differential reddening explained above. The morphology of the CMD is consistent with what was found in previous works \citep{piotto2002,cote1997}, confirming that the cluster does indeed contain an extended blue HB. However, our work is the first to clearly reach beyond the MS turn-off level, and to confirm the presence of blue HB stars at least as faint as the turn-off point in $V$. Conversely, we do not identify a prominent red HB component, contrary to what is commonly seen at such moderately high metallicities~-- thus confirming that the cluster is indeed a second-parameter object. 

\begin{figure}[t]
\begin{center}
 \resizebox{8cm}{!}{\includegraphics{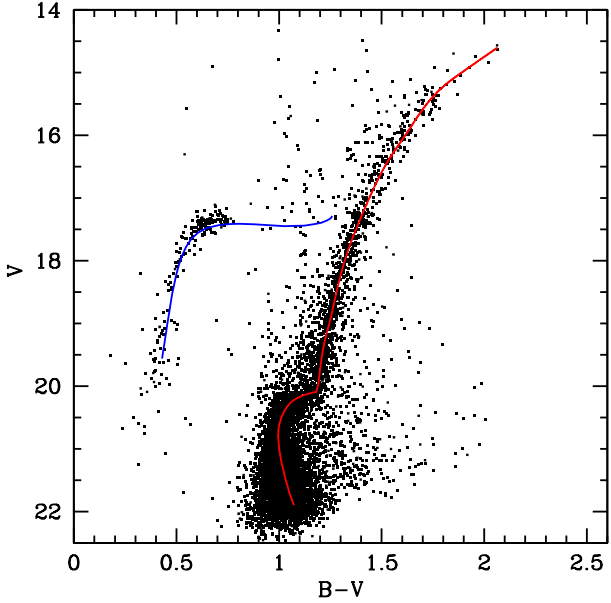}}
 \caption{CMD showing the ZAHB (\textit{blue}) and the fit to the fiducial points of the RGB along with the MS ridgeline (\textit{red}).}
 \label{hb2}
 \end{center}
\end{figure}

Based on our CMD, we estimate the metallicity of M14 by determining the photometric indices from \citet{f99}. In order to use their equations, we first need to precisely determine the cluster's ridgeline, and most importantly the HB level, since any changes in this level will affect our final results. To determine the ridgeline, we follow the method described in (e.g.) \citet{mz1} and \citet{Stet2005}: we first determine our fiducial points for the cluster by visual inspection, and then fit a cubic spline to the fiducial points of the RGB thus derived. The spline fit to the fiducial points of the RGB, along with the mean ridgeline of the MS, can be observed in Figure~\ref{hb2}. The fiducial points derived for the cluster are also given in Table~\ref{tab:ridge}.

\begin{deluxetable}{ll}
\tablewidth{0pc} 
%\tabletypesize{\scriptsize}
\tablecaption{M14 Ridgeline}
\tablehead{
\colhead{$V$} & 
\colhead{$\bv$} 
}
\startdata
14.611  &  2.062  \\ 
14.800  &  1.978  \\
14.999  &  1.889  \\ 
15.210  &  1.805  \\ 
15.499  &  1.720  \\ 
15.740  &  1.668  \\ 
15.999  &  1.615  \\ 
16.240  &  1.565  \\ 
16.492  &  1.516  \\ 
16.734  &  1.478  \\ 
16.993  &  1.441  \\ 
17.244  &  1.408  \\ 
17.494  &  1.376  \\ 
17.735  &  1.348  \\ 
17.995  &  1.321  \\ 
18.229  &  1.300  \\ 
18.495  &  1.281  \\ 
18.754  &  1.263  \\ 
18.995  &  1.246  \\ 
19.214  &  1.228  \\ 
19.571  &  1.207  \\ 
19.727  &  1.200  \\ 
19.978  &  1.191  \\ 
20.000  &  1.190  \\ 
20.050  &  1.185  \\ 
20.100  &  1.180  \\ 
20.100  &  1.167  \\ 
20.150  &  1.114  \\ 
20.200  &  1.083  \\ 
20.250  &  1.060  \\ 
20.300  &  1.046  \\ 
20.350  &  1.035  \\ 
20.400  &  1.025  \\ 
20.450  &  1.018  \\ 
20.500  &  1.011  \\ 
20.550  &  1.007  \\ 
20.600  &  1.002  \\ 
20.650  &  0.999  \\ 
20.700  &  0.997  \\ 
20.750  &  0.996  \\ 
20.800  &  0.995  \\ 
20.850  &  0.996  \\ 
20.900  &  0.998  \\ 
20.950  &  0.999  \\ 
21.000  &  1.000  \\ 
21.100  &  1.005  \\ 
21.200  &  1.011  \\ 
21.300  &  1.019  \\ 
21.400  &  1.025  \\ 
21.500  &  1.033  \\ 
21.600  &  1.042  \\ 
21.700  &  1.049  \\ 
21.800  &  1.061  \\ 
21.900  &  1.073  
\enddata
\label{tab:ridge}
\end{deluxetable}

\begin{figure}[t]
\begin{center}
 \resizebox{8cm}{!}{\includegraphics{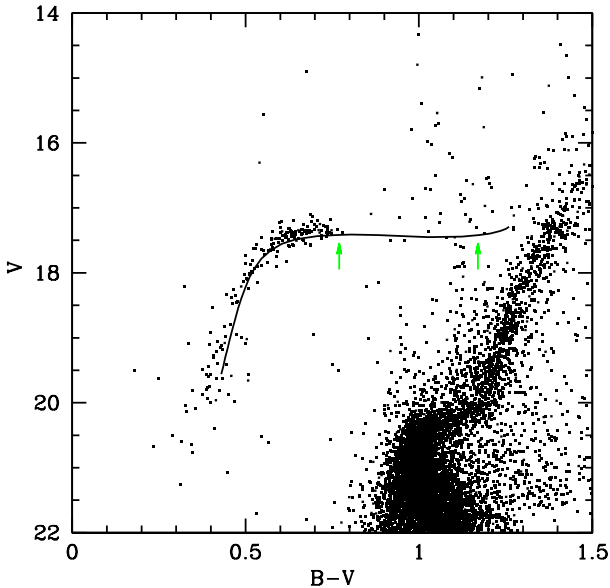}}
 \caption{Position of the ZAHB models from \citet{vdb2006}, for a chemical composition ${\rm [Fe/H]} = -1.41$, $[\alpha/{\rm Fe}]=0.3$, superimposed on the data. The green arrows mark the limits of the region used to determine the mean magnitude of the ZAHB (see text for further details).}
 \label{hb1}
 \end{center}
\end{figure}

\citet{f99} use as the HB level the empirical zero-age HB (ZAHB), which is defined by them as the magnitude of the lower envelope of the observed HB distribution in the region with $0.2 < (B-V)_{0} < 0.6$. This amounts to the effectively ``horizontal'' part of the HB in the CMD, where unfortunately M14 does not contain many stars. Moreover, this region of the CMD may also suffer from uncertainties in our procedure to perform statistical decontamination. In view of these limitations, we decided to use a theoretical ZAHB from \citet{vdb2006} for a chemical composition ${\rm [Fe/H]} = -1.41$ (see below), and an enhancement of the $\alpha$-elements given by $[\alpha/{\rm Fe}]=0.3$, as typically found among Galactic GC stars \citep[e.g.][]{1996Carney,2004Sneden}. We use $E(\bv)=0.57$ (\S\ref{sec:red}), and vertically shift the theoretical ZAHB to match the lower envelope of our observed blue HB distribution (Fig.~\ref{hb1}). The HB level is then determined as the mean magnitude of the ZAHB in the region $0.77 < (B-V) < 1.17$. This gives $V_{\rm ZAHB}=17.45$. 

Using our mean ridgeline, we derive the RGB parameters $(B-V)_{0,g}$, $\Delta V_{1.1}$, $\Delta V_{1.2}$, $\Delta V_{1.4}$, $S_{2.0}$, and $S_{2.5}$. For the first four among these indices we need to correct our observed colors by extinction, which was done by assuming once again $E(\bv)=0.57$. The RGB parameters and the respective metallicities derived on the basis of each index are listed in Table~\ref{tmeta1}. A straight average over the derived values gives ${\rm [Fe/H]}_{\rm CG97} = -1.15 \pm 0.06$ in the \citet{cg1} scale. This corresponds, following Eq.~7 in \citeauthor{cg1}, to a value ${\rm [Fe/H]}_{\rm ZW84} = -1.38 \pm 0.07$ in the \citet{zw84} scale. This is close to the value listed in the 2003 edition of the \citet{h96} catalog, namely ${\rm [Fe/H]} = -1.39$, but slightly lower than the value provided in the 2010 edition of this catalog \citep{wh10}, namely ${\rm [Fe/H]} = -1.28$. The latter value is, however, in the new UVES spectroscopic scale of \citet{cg2}. Using the transformation relations given in \citeauthor{cg2}, our metallicity value translates into ${\rm [Fe/H]}_{\rm UVES} = -1.27$, in excellent agreement with \citet{wh10}. Indeed, we obtain a mean value ${\rm [Fe/H]}_{\rm UVES} = -1.28\pm 0.08$, based on the different photometric indicators (Table~\ref{tmeta1}).

The indices $S_{2.0}$ and $S_{2.5}$ are independent of the assumed reddening of the cluster, and thus can be used as a check of the adopted reddening value and derived metallicity. From Table~\ref{tmeta1} we can see that the values of [Fe/H] obtained with these two slopes do not differ from those obtained with reddening-dependent indices. Indeed, based on $S_{2.0}$ and $S_{2.5}$ we derive $(\bv)_{0,g}=0.84$ and $(\bv)_{0,g}=0.83$, respectively. It is thus possible to conclude that the assumed value of $E(\bv)=0.57$, which implies $(\bv)_{0,g} = 0.81$,  must be close to the correct reddening for M14, to within $\approx 0.02$~mag in $E(\bv)$.

\begin{deluxetable}{cccc}
\tablewidth{0pc} 
%\tabletypesize{\scriptsize}
\tablecaption{Photometric Metallicity Indicators in M14\tablenotemark{a}}
\tablehead{
\colhead{Parameter} & 
\colhead{${\rm [Fe/H]}_{\rm CG97}$} & 
\colhead{${\rm [Fe/H]}_{\rm ZW84}$} & 
\colhead{${\rm [Fe/H]}_{\rm UVES}$}
}
\startdata
$(\bv)_{0,g} = 0.81$    & $-1.27$ & $-1.51$ & $-1.42$ \\
$\Delta V_{1.1} = 1.72$ & $-1.10$ & $-1.32$ & $-1.20$ \\
$\Delta V_{1.2} = 2.13$ & $-1.14$ & $-1.36$ & $-1.25$ \\
$\Delta V_{1.4} = 2.63$ & $-1.17$ & $-1.40$ & $-1.29$ \\
$S_{2.5} = 4.69$        & $-1.15$ & $-1.37$ & $-1.26$ \\
$S_{2.0} = 5.68$        & $-1.12$ & $-1.34$ & $-1.23$ \\
Mean & $-1.15\pm 0.06$ & $-1.38 \pm 0.07$  &  $-1.28 \pm 0.08$
\enddata
\tablenotetext{a}{Based on the \citet{f99} calibrations. }
\label{tmeta1}
\end{deluxetable}

%\begin{table}[t]
%\begin{center}
%\footnotesize
%\caption{\footnotesize{Mean Fiducial Points for NGC~5286}}
%\begin{tabular}{lc}
%\tableline\tableline
%{\it{V}} & (\bv) \\
%\tableline
%\tableline
%\multicolumn{2}{c}{MS + SGB\tablenotemark{a} + RGB}\\
%\tableline
%$13.602$........................... & $1.793$ \\
%$13.636$........................... & $1.692$ \\
%$13.785$........................... & $1.602$ \\
%$13.933$........................... & $1.537$ \\
%$14.213$........................... & $1.441$ \\
%$14.464$........................... & $1.356$ \\
%$14.771$........................... & $1.268$ \\
%$15.135$........................... & $1.197$ \\
%$15.470$........................... & $1.136$ \\
%$15.819$........................... & $1.090$ \\
%$16.210$........................... & $1.048$ \\
%$16.643$........................... & $1.002$ \\
%$17.076$........................... & $0.966$ \\
%$17.495$........................... & $0.941$ \\
%$17.886$........................... & $0.917$ \\
%$18.263$........................... & $0.899$ \\
%$18.696$........................... & $0.880$ \\
%$19.129$........................... & $0.842$ \\
%$19.297$........................... & $0.821$ \\
%\tableline
%\multicolumn{2}{c}{Blue HB}\\
%\tableline
%$16.671$........................... & $ 0.378$  \\  
%$16.769$........................... & $ 0.303$  \\  
%\tableline
%\tablenotetext{a}{ SGB = subgiant branch.} 

%\end{tabular}
%\label{tablaridge}
%\end{center}
%\end{table}       

\subsection{Horizontal Branch Morphology}  \label{sec:hbmorp}

As mentioned above, NGC~6402 presents an unusually blue HB for its metallicity. We attempt, in the following, to describe its HB morphology through widely used parameters, such as the Lee-Zinn parameter $\mathcal{L}$ \citep{1986Zinn,1990Lee2,1990Lee3} and the Buonnano parameter $\mathcal{P}_{\mathcal{HB}}$ \citep{1993Buonanno,1997Buonanno}. These are defined as follows:

\begin{eqnarray}
\mathcal{L} & = & \mathcal{(B-R)/(B+V+R)},\\
\mathcal{P}_{\mathcal{HB}} & = & \mathcal{(B{\rm 2}-R)/(B+V+R)}, 
\end{eqnarray}

\noindent where $\mathcal{(B{\rm 2}}$ represents the number of blue-HB stars bluer than $(B-V)_{0} < -0.02$ and $\mathcal{B,\, R,\, V}$ represent the number of blue, red, and variable stars in the HB (RR Lyrae), respectively. Two other parameters are also presented in this study, both defined by \citet{1991Preston}: $(B-V)_{W}$, or the mean unreddened color of blue HB stars with $-0.02 < (B-V)_{0} < 0.18$, and $B_{W}/\mathcal{B}$, defined as the number of stars in the given range in color, normalised by the number of blue HB stars with $(B-V)_{0}<0.18$.

\begin{figure}[t]
\begin{center}
 \resizebox{8cm}{!}{\includegraphics{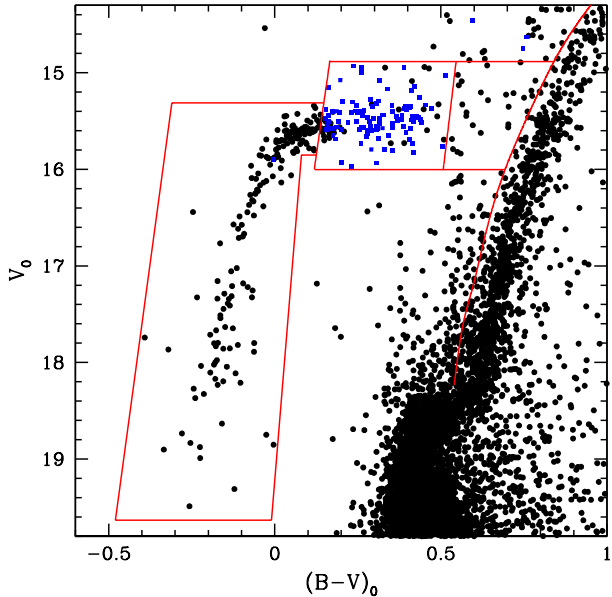}}
 \caption{Selection boxes, in the M14 CMD, for (from right to left) the red, variable, and blue HB stars, respectively. Identified RR Lyrae variables (from Paper~II) are shown in blue.}
 \label{HBmorp}
 \end{center}
\end{figure}

Blue, red and variable stars in the HB were selected following the boundary regions shown in Figure~\ref{HBmorp}. The criteria adopted can be summarized as follows. 

\begin{itemize}
\item $\mathcal{R} = $ number of red HB stars: The red HB region is limited, at its red end, by the RGB ridgeline minus 0.08~mag, corresponding to the (somewhat arbitrary) blue limit that was set to select stars in the study of the RGB bump in \S\ref{sec:rgb}. The faint limit is set as the lower envelope of the observed RR Lyrae distribution, whereas the bright limit was adopted at $V_{\rm ZAHB}-0.8$, as needed in order to separate HB stars from asymptotic giant branch stars \citep{1994Buonanno}. Finally the blue limit is set by the red limit of the RR Lyrae distribution.
\item $\mathcal{V} = $ number of RR Lyrae stars: This quantity is taken from Paper~II, corresponding to all RR Lyrae found within the same magnitude limits as defined for red HB stars.
\item $\mathcal{B} = $ number of stars contained in the fainter/bluer selection box shown in Figure~\ref{HBmorp}: The limits of this selection box are set by hand, in order to include likely faint blue HB stars affected by large photometric errors.  

\end{itemize}

We note that ``constant'' stars lying within the $V$ selection box are defined as red or blue HB stars, depending on its distance to either one of these groups.

The number of stars selected for each region as explained above, and the resulting parameters of the morphology of the HB are presented in Table~\ref{hbtab}.

The computed value of $\mathcal{L}$ is much lower than the one found in the literature for this cluster \citep[$\mathcal{L}=0.65$, see][and references therein]{2009Catelan}. The uncertainty is mainly due to the problems involved in accurately evaluating $\mathcal{R}$, given that it is strongly affected by the uncertainties in the CMD decontamination process. Note, in this sense, that the original $\mathcal{L}$ value for this cluster was based on the assumption that $\mathcal{R} = 0$ \citep{1994Lee}.

 % PROBABLY PUT $\mathcal{L}$ vs $[Fe/H]$ PLANE TO ILLUSTRATE THE OOSTERHOFF INTERMIDIATE CLASSIFICATION

\begin{table}[h!]
\begin{center}
\footnotesize
\caption{\footnotesize{HB Morphology Parameters for NGC~6402.}}
\begin{tabular}{lcc}
\tableline\tableline
Parameter & Value \\
\tableline
$\mathcal{B:V:R}$	& $191 : 106 : 39 $ \\
$\mathcal{B/(B+R)}$ & $ 0.83 \pm 0.02$ \\
$\mathcal{(B-R)/(B+V+R)}$ & $0.45 \pm 0.03$ \\
$\mathcal{(B{\rm 2}-R)/(B+V+R)}$ & $0.14 \pm 0.03$ \\
$(B-V)_W$ &  $0.078 \pm 0.054$ \\
$B_W/\mathcal{B}$	&  $0.54 \pm 0.14$ \\
\tableline
\end{tabular}
\label{hbtab}
\end{center}
\end{table}

%\begin{table}[h!]
%\begin{center}
%\footnotesize
%\caption{\footnotesize{HB Morphology Parameters for NGC~5286}}
%\begin{tabular}{lcc}
%\tableline\tableline
%Parameter & Value \\
%\tableline
%$\mathcal{B:V:R}$	& $0.764 : 0.338 : 0.027$ \\
%$\mathcal{B/(B+R)}$	& $0.973 \pm 0.03$ \\
%$\mathcal{(B-R)/(B+V+R)}$	& $0.739 \pm 0.02$ \\
%$\mathcal{(B{\rm 2}-R)/(B+V+R)}$	& $0.342 \pm 0.02$ \\
%$(\bv)_W$	&  $0.074 \pm 0.005$ \\
%$B_W/B$	&  $0.523 \pm 0.05$ \\
%\tableline
%\end{tabular}
%\label{hbtab}
%\end{center}
%\end{table} 

    %\begin{figure}[t]
      %\plotone{bump-analysis.eps}
    %  \caption{\footnotesize{
%{\em Top}: differential RGB luminosity function for NGC~5286 ({\em solid line}). A power-law fit to the data is shown as a {\em dash-dotted gray line}. {\em Middle}: the residuals around the power-law fit are shown as a {\em solid line}. The {\em dashed lines} indicate the Poisson $3-\sigma$ level, as computed based on the derived power-law fit. The gray line shows the Gaussian fit over a $\pm 0.3$~mag region around the candidate RGB bump peak. {\em Bottom}: cumulative luminosity function, indicating the break expected if the bump is present at the indicated location. All three panels consistently indicate that the RGB bump is located at the position marked by the arrow. 
	%  }}
	  %\label{historgb}
      %\end{figure} 

\subsection{Red Giant Branch}  \label{sec:rgb}
\begin{figure}[t]
\begin{center}
 \resizebox{8cm}{!}{\includegraphics{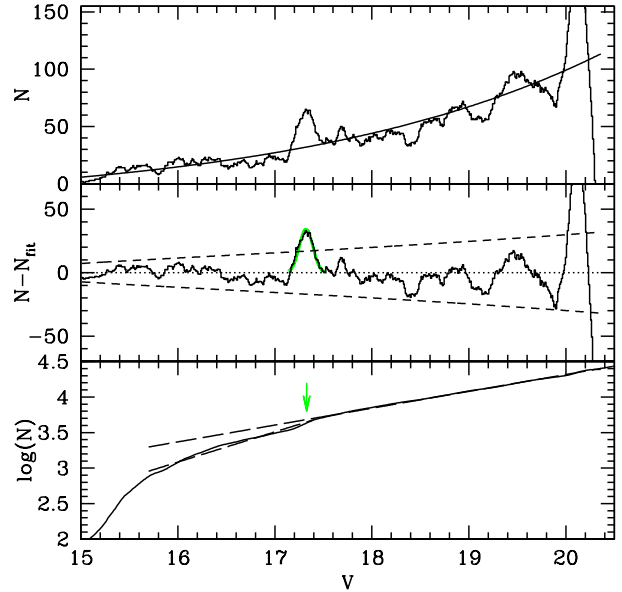}}
 \caption{({\em top}) Smoothed RGB LF. A power-law fit to the LF is overplotted ({\em black line}). ({\em middle}) Differential RGB LF, $N-N_{\rm fit}$, along with the expected $3-\sigma$ Poisson fluctuation level ({\em long-dashed black line}). A Gaussian fit to the suspected feature associated to the RGB bump is plotted in {\em green}. ({\em bottom}) Integrated RGB LF. The green arrow marks the derived position of the RGB bump, at $V_{\rm bump}=17.32\pm0.06$~mag.}
 \label{rgb_bump}
 \end{center}
\end{figure}

One of the most important phases in the evolution of an RGB star is the so-called RGB ``bump.'' This corresponds to a momentary reversal in the evolutionary path of the star, which takes place when its H-burning shell encounters the chemical composition discontinuity left behind by the maximum inward penetration of the giant's convective envelope \citep[see][for a review and extensive references]{mc07}. In order to determine the position of the RGB bump in M14, we follow the same procedure as applied by \citet{mz1} in the case of NGC~5286. The smoothed RGB luminosity function (LF; Fig.~\ref{rgb_bump}, {\em upper panel}) already reveals the presence of a candidate bump at $V\sim 17.3$. The significance of the detection is confirmed by subtracting the expected number of stars $N_{\rm fit}$ given by a power-law fit to the LF from the number of stars in each bin, and comparing this result to the expected $3-\sigma$ Poisson fluctuation level. As can be seen in the {\em middle panel} of Figure~\ref{rgb_bump}, the suspected bump clearly stands out as the most prominent feature (the higher peak at $V \sim 20$ is due to MS stars). A Gaussian fit to the data around this feature yields $V_{\rm bump}=17.321\pm0.003$,  with $\sigma_{V,{\rm bump}}=0.070\pm0.002$. We accordingly estimate the final position of the bump as $V_{\rm bump}=17.32\pm0.06$, which results from a compromise between the formal error of the position and the width of the Gaussian \citep[see also][]{mz1}. The location of the bump is also confirmed by the integrated RGB LF, as shown in the {\em bottom panel} of Figure~\ref{rgb_bump}.

\begin{figure}[t]
\begin{center}
 \resizebox{8cm}{!}{\includegraphics{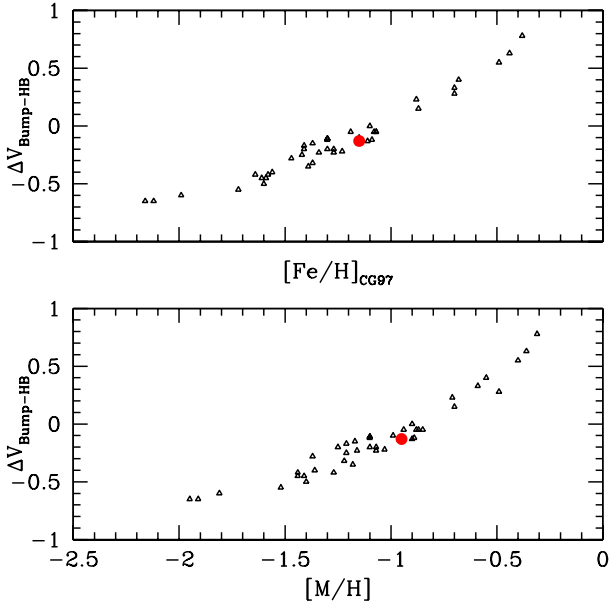}}
 \caption{The position of M14 in the $\Delta V^{\rm bump}_{\rm HB}$ vs. metallicity plane ({\em red circle}) is compared against those of other GCs, using data from \citet{f99}.}
 \label{bumpf99}
 \end{center}
\end{figure}

The derived position of the RGB bump and ZAHB imply a $\Delta V^{\rm bump}_{\rm HB} = -0.13$~mag. As well known, this quantity is also strongly sensitive to the metallicity, and thus in Figure~\ref{bumpf99} we present $\Delta V^{\rm bump}_{\rm HB}$ as a function of [Fe/H] ({\em upper panel}) and [M/H] ({\em lower panel}), using the data from \citet{f99}. Clearly, the position of the RGB bump in M14 is consistent with our derived metallicity for the cluster, again lending support to our adopted reddening value.

\section{Age}\label{sec:age}
 
We make an attempt in determining the age of M14 by comparing to the GC M5 \citep{1996Sandquist}, using two methods.

Firstly we use the so-called $\Delta V$ method, based on the magnitude difference between the HB and TO levels. For this we follow \citet{1996Chaboyer}, using their equation (2) and assuming $M_{V}({\rm RR}) = 0.20\, {\rm [Fe/H]}+0.98$. The latter is assumed as it corresponds to the closest match to the absolute magnitudes of RR Lyrae obtained from the recent calibration by \citet{2008Catelan_cortes}. The HB level of M14 is determined, as described in \citet{1996Chaboyer}, by using $V({\rm HB})=V_{{\rm ZAHB}}-0.05\,{\rm [Fe/H]}-0.2$ (where ${\rm [Fe/H]}=-1.38$), which yields $V({\rm HB})=17.32\pm0.01$. The TO level is determined by fitting a parabola to a small region of the MS near the TO point, thus obtaining $V_{\rm TO}=20.81\pm0.02$ and $(B-V)_{\rm TO}=0.995\pm0.030$. Then, the corresponding $\Delta V$ value for M14 is $\Delta V^{{\rm TO}}_{{\rm HB}}=3.49\pm0.02$, while for M5, $\Delta V^{\rm TO}_{\rm HB}=3.47\pm0.06$ \citep[from][]{1996Sandquist}. This implies that M14 is $\sim 0.3\pm1.0$~Gyr older than M5.

\begin{figure}[t]
\begin{center}
 \resizebox{8cm}{!}{\includegraphics{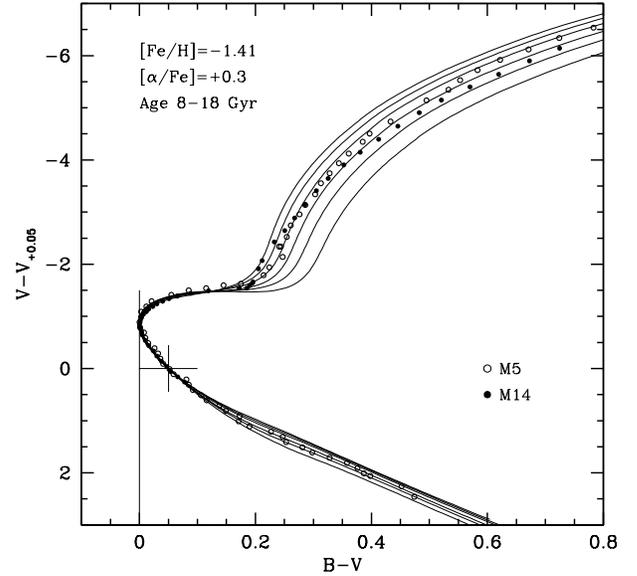}}
 \caption{Comparison of the ridgelines of M14 (\textit{filled circles}) and M5 (\textit{open circles}) with the theoretical isochrones of \citet{vdb2006} for ${\rm [Fe/H]}=-1.41$, $[\alpha/{\rm Fe}]=0.3$, and ages in the range $[8,18]$~Gyr, in 2~Gyr steps (\textit{solid lines}). The models and data are registered as explained in the text.}  
 \label{vdb}
 \end{center}
\end{figure}

Figure~\ref{vdb} shows the fiducial points of M14 as determined by us and those of M5 taken from \citet{1996Sandquist}. In addition, \citet{vdb2006} isochrones for ${\rm [Fe/H]}=-1.41$, $[\alpha/{\rm Fe}]=0.3$, and ages ranging from 8 to 18~Gyr are also shown. For comparison with the models, we chose the closest metallicity to the \cite{zw84} value determined for M14.

Following \citet{1999Stetson}, the model isochrones and ridgelines are translated horizontally to match each other's turnoff colors, and then shifted vertically to register the point of the upper MS that is 0.05~mag redder than the turnoff. The age is derived using Equation~(1) in \citet{1991Vdb}, which requires to measure the color difference between the RGBs. \citet{1999Stetson} measure this difference at the point that lies exactly at $V-V_{+0.05}=-3$, whereas \citet{1991Vdb} use the value determined at $V-V_{+0.05}=-2.5$; however, the latter authors recommend to perform a star-by-star statistical analysis over 1-2 magnitudes along the lower RGB, in order to more reliably measure the mean color difference between the RGBs. It is clear from Figure~\ref{vdb} that for our comparison, a value derived from this method will strongly depend on the point where the color difference is evaluated, thus we make no attempt on deriving an age difference. However, the data appear to suggest that M14 is slightly older than M5, and that the age difference could be as high as 2~Gyr. This is consistent with our previous estimate, based on the vertical method. A slightly older M14, compared to M5, would go in the right sense to help explain the former's bluer HB morphology. However, our data are not sufficient to reach definite conclusions in this regard, and so we are unable to exclude the possibility that M14 and M5 have virtually the same age, to within the errors. 
 
\section{Summary} \label{sec:summ}

We have obtained $BVI$ CCD photometry for the GC \objectname{M14}. A CMD of unprecedented depth is constructed for the cluster, based on which we derive several key properties.

The CMD is found to be affected by contamination by field stars, as well as the presence of differential reddening across the face of the cluster. The latter is estimated to be present at the level of $\Delta E(\bv) \approx 0.17$~mag. After statistically subtracting the field stars, a reddening map is built, and the photometry of M14 is finally corrected for the presence of differential reddening.

RGB photometric parameters are derived as a means to obtain the cluster's metallicity, implying ${\rm [Fe/H]} = -1.38\pm 0.07$~dex in the \citet{zw84} metallicity scale, and a mean value ${\rm [Fe/H]} = -1.28\pm 0.08$~dex on the more recent UVES scale, in excellent agreement with the value listed in the \citet{wh10} catalog. We detect the RGB bump at $V_{\rm bump} = 17.32 \pm 0.06$~mag, and its position is also fully consistent with the derived metallicity. 

The derived photometric parameters allow us to confirm that the absolute reddening of the cluster must be close to $E(B-V)=0.57$, to within about 0.02~mag. The latter is significantly higher than implied by the dust maps of \citeauthor{1982BHM} or \citeauthor{1998SchM}.

HB morphology parameters are also computed, with the main difference between our results and those of previous works arising from the uncertainty in the number of stars belonging to the red part of the HB, caused by the decontamination process.

The so-called $\Delta V$ method is used to determine relative ages between M14 and M5, suggesting that M14 is only about $0.3\pm1.0$~Gyr older than M5~-- a result qualitatively also supported by a comparison with model isochrones and the ``horizontal'' age method. Therefore, within the uncertainties, M14 and M5 have roughly the same age, although the possibility that M14 may be slightly older (by up to $\approx 2$~Gyr) cannot be ruled out.

\acknowledgments
We thank an anonymous referee for several useful suggestions that 
helped us improve the presentation of our results.
Support for M.C. and C.C.P. is provided by the Chilean Ministry for the
Economy, Development, and Tourism's Programa Iniciativa Cient\'{i}fica 
Milenio through grant P07-021-F, awarded to The Milky Way Millennium 
Nucleus; by the BASAL Center for Astrophysics and Associated Technologies 
(PFB-06); by Proyecto Fondecyt Regular \#1110326; and by Proyecto Anillos 
ACT-86.

\bibliography{ref}

\end{document}